\documentclass[journal]{IEEEtran}
\usepackage[linesnumbered,ruled]{algorithm2e}
\usepackage[usenames, dvipsnames]{color}
\usepackage{graphicx,amssymb,mathrsfs,amsmath,array}
\usepackage{subfigure}
\usepackage{multirow}
\usepackage{booktabs}
\usepackage{xcolor}
\usepackage{cases}
\usepackage{epstopdf}
\graphicspath{{figures/}}

\usepackage[colorlinks,urlcolor=blue,driverfallback=dvipdfm]{hyperref}

\usepackage{algpseudocode}
\usepackage{cite}

\newcommand{\norm}[1]{\lVert#1\rVert}

\DeclareMathOperator{\F}{F}
\DeclareMathOperator{\T}{T}

\begin{document}
\title{Low-Rank Representations Meets Deep Unfolding: A Generalized and Interpretable Network for Hyperspectral Anomaly Detection}

\author{Chenyu Li,~\IEEEmembership{Student Member,~IEEE,}
        Bing Zhang,~\IEEEmembership{Fellow,~IEEE,} 
        Danfeng Hong,~\IEEEmembership{Senior Member,~IEEE,}
        Jing Yao,~\IEEEmembership{Member,~IEEE,}     
        and Jocelyn Chanussot,~\IEEEmembership{Fellow,~IEEE}

\thanks{This work was supported by the National Natural Science Foundation of China under Grant 42241109 and Grant 42271350 and by the MIAI@Grenoble Alpes (ANR-19-P3IA-0003). (\emph{Corresponding author: Bing Zhang})}
\thanks{C. Li is with the Aerospace Information Research Institute, Chinese Academy of Sciences, 100094 Beijing, China, and also the School of Mathematics and Statistics, Southeast University, 211189 Nanjing, China. (e-mail: lichenyu@seu.edu.cn)}
\thanks{B. Zhang is with the Aerospace Information Research Institute, Chinese Academy of Sciences, 100094 Beijing, China, and also the College of Resources and Environment, University of Chinese Academy of Sciences, Beijing 100049, China. (e-mail: zb@radi.ac.cn)}
\thanks{D. Hong and J. Yao are with the Aerospace Information Research Institute, Chinese Academy of Sciences, 100094 Beijing, China. (e-mail: hongdf@aircas.ac.cn, yaojing@aircas.ac.cn)}
\thanks{J. Chanussot is with Univ. Grenoble Alpes, Inria, CNRS, Grenoble INP, LJK, Grenoble, 38000, France, and also with the Aerospace Information Research Institute, Chinese Academy of Sciences, 100094 Beijing, China. (e-mail: jocelyn.chanussot@inria.fr)}
}

\markboth{Submission to IEEE TGRS ,~Vol.~XX, No.~XX, ~XXXX,~2023}
{Shell \MakeLowercase{\textit{et al.}}: }
\maketitle

\begin{abstract}
Current hyperspectral anomaly detection (HAD) benchmark datasets suffer from low resolution, simple background, and small size of the detection data.  These factors also limit the performance of the well-known low-rank representation (LRR) models in terms of robustness on the separation of background and target features and the reliance on manual parameter selection. To this end, we build a new set of HAD benchmark datasets for improving the robustness of the HAD algorithm in complex scenarios, AIR-HAD for short. Accordingly, we propose a generalized and interpretable HAD network by deeply unfolding a dictionary-learnable LLR model, named LRR-Net$^+$, which is capable of spectrally decoupling the background structure and object properties in a more generalized fashion and eliminating the bias introduced by vital interference targets concurrently. In addition, LRR-Net$^+$ integrates the solution process of the Alternating Direction Method of Multipliers (ADMM) optimizer with the deep network, guiding its search process and imparting a level of interpretability to parameter optimization.
Additionally, the integration of physical models with DL techniques eliminates the need for manual parameter tuning. The manually tuned parameters are seamlessly transformed into trainable parameters for deep neural networks, facilitating a more efficient and automated optimization process. Extensive experiments conducted on the AIR-HAD dataset show the superiority of our LRR-Net$^+$ in terms of detection performance and generalization ability, compared to top-performing rivals. Furthermore, the compilable codes and our AIR-HAD benchmark datasets in this paper will be made available freely and openly at \url{https://sites.google.com/view/danfeng-hong}.
\end{abstract}


\section{Introduction}
\IEEEPARstart{T}{he} swift advancement of space technology and the continuous growth of remote sensing (RS) data sources for satellite and airborne sensors has promoted the extensive use of RS technology. A hyperspectral image (HSI) has dozens of spectral bands, with a wavelength range possibly from infrared to ultraviolet light \cite{1985Imaging}. In contrast to RGB images with only three spectral bands, HSI contains rich spectral information showing features of objects hidden in the spectral domain, applied to camouflage recognition, fine agriculture, mineral recognition, change detection, and land cover classification \cite{hong2024spectralgpt}. Among these applications, hyperspectral anomaly detection (HAD) is an important task of hyperspectral image processing \cite{974724}, which assesses whether a testing point demonstrates any atypically, enabling the differentiation between background and abnormal targets within the observed data. Due to the limitations of the data set, HAD task is usually unsupervised.
\begin{figure}[!t]
	  \centering	 
           \includegraphics[width=0.45\textwidth]{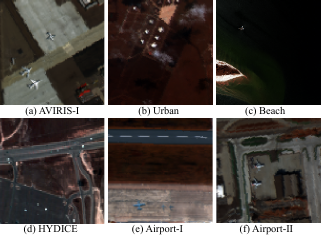}
        \caption{Pseudo-color images of different public data sets in hyperspectral anomaly detection.}
\label{fig1}
\end{figure}

\begin{table*}[!t]
\centering
	\caption{Comparison of the public datasets for the HAD task.}
    \scalebox{1.0}{
	\begin{tabular}{c|c|c|c|c|c|c|c}
		\toprule
		\multirow{1}{*} Images & Captured place & Spatial Resolution & Spectral Resolution & Band  & Image size & Sensor & Flight time\\
    &   & (m) & (nm) &    & (pixels) &   &  \\
		\hline\hline
Airport & Los Angeles & 7.1 & 10.0 &  205 & 100 $\times$ 100 & AVIRIS & 11/2011  \\
        
                 & Los Angeles & 7.1 & 10.0 & 205 & 100 $\times$ 100 & AVIRIS & 11/2011  \\ 
  
                 & Los Angeles & 7.1 & 10.0 & 205 & 100 $\times$ 100 & AVIRIS & 11/2011  \\

                 & Gulfport & 3.4 & 10.0 & 191 & 100 $\times$ 100 & AVIRIS & 7/2010  \\        
\hline \hline
        Beach & Cat Island & 17.2 & 10.0 & 188 & 150 $\times$ 150 & AVIRIS & 9/2010  \\
  
               & San Diego & 7.5 & 10.0 & 193 & 100 $\times$ 100 & AVIRIS & 11/2011  \\ 
  
               & Bay Champagne & 4.4 & 10.0 & 188 & 100 $\times$ 100& AVIRIS & 7/2010 \\

               & Pavia & 1.3 & 10.0 & 102 & 150 $\times$ 150 & ROSIS-03 & Unknown  \\  
\hline \hline
        Urban  & Texas Coast & 17.2 & 10.0 & 204 & 100 $\times$ 100 & AVIRIS & 8/2010  \\
  
               & Texas Coast & 17.2 & 10.0 & 207 & 100 $\times$ 100 & AVIRIS & 8/2010  \\ 
  
               & Gainesville & 3.5 & 10.0 & 191 & 100 $\times$ 100 & AVIRIS & 9/2010  \\

               & Los Angeles & 7.1 & 10.0 & 205 & 100 $\times$ 100 & AVIRIS & 11/2011  \\  

               & Los Angeles & 7.1 & 10.0 & 205 & 100 $\times$ 100 & AVIRIS & 11/2011  \\ 
        \hline\hline
        HYDICE & Urban & 1.56 & 10.0 & 175 & 80 $\times$ 100 & HYDICE & 8/1995  \\    
\hline\hline
        San Diego & California & 3.5 & 10.0 & 186 & 100 $\times$ 100 & AVIRIS & 8/1995   \\ 
		\bottomrule[1.5pt]
	\end{tabular}
 }
    \label{Tab1}
\end{table*}

Excellent and challenging publicly available datasets significantly contribute to verifying the effectiveness of HAD algorithms, and can also indirectly reflect the stability and generalization of the algorithm \cite{2010Learning, 8550778, li2014collaborative}. Benefiting from the availability of the data collected by the Airborne Visible/Infrared Imaging Spectrometer (AVIRIS) and the Hyperspectral Digital Imagery Collection Experiment (HYDICE) sensors \cite{Cheng2020TGRS_gtvlrr,9133150}. The public availability of the test data sets has provided outstanding contributions in HAD algorithm research, mainly including San Diego, Airport-Beach-Urban (ABU), and HYDICE, and some parameters were compiled in Tab. \ref{Tab1}. Nevertheless, the existing data sets have several drawbacks that result in their validation capabilities remaining limited in terms of robustness and generality of HAD. 

\begin{itemize}
    \item \textbf {The resolution of remote sensing images needs to be further improved.} The existing public test datasets were acquired by shooting ten years ago, with low spatial or spectral resolution, limiting the efficacy of HAD algorithms. In Tab. \ref{Tab1}, the spatial resolution is mainly from meter level to more than ten meters level. Meanwhile, the spectral resolution is also below ten nanometers.
    \item \textbf {The type of geographical features need to be enriched.} There is an insufficient variety of non-abnormal features, and at the same time, there are large spectral differences between abnormal pixels and background pixels, which poses no significant challenges for the HAD algorithm and may mildly impact the model's generalization ability.
    \item \textbf {The size of the dataset needs to be increased.} The size of existing datasets usually does not exceed 100$\times$100 pixels, and some advanced algorithms can achieve superior performance on small-scale datasets but have poor experimental results on a large range of datasets. To more comprehensively assess the effectiveness of the HAD algorithm, it is necessary to utilize a relatively large-scale test dataset.  
\end{itemize}

Conventional methods for HAD typically rely on modeling the differences in mathematical statistics or spatial distribution features to identify anomalies targets \cite{reed1990adaptive}. However, these methods, which assume statistical distribution patterns, often struggle to detect complex feature types distributions as remote sensing (RS) application requirements become more challenging \cite{8519775}. On the other hand, hyperspectral data exhibits redundancy and high correlation, making it suitable for leveraging the low-rank representation (LRR) technique to capture global data information \cite{Liu2013TPAMI, chen2013SPIE}. The LRR model hypothesizes that data points reside in multiple underlying subspaces, enabling each data vector to be represented as the lowest-rank linear weighted total with a shared dictionary \cite{niu2016LRLD_RS}. To overcome the challenges associated with hyperspectral image anomaly detection, a novel technique known as the Abundance and Dictionary-based Low-rank Representation (ADLR) method was recently proposed in the literature \cite{8335758}. Recognizing the significance of local geometrical and spatial information in discriminating between interference and targets, the LRR model was extended with the incorporation of graph regularization and total variation (TV) regularization for HSI anomaly detection \cite{Cheng2020TGRS_gtvlrr}. Following this line of research, scholars have discovered that low-rank regularization effectively captures global structural information, while sparse regularization excels at representing local information \cite{xu2016TGRS_LRSAR}. Taking inspiration from this insight, we devised a novel approach known as the Low-Rank and Sparse Representation (LRASR) method for the HAD task. The method uses a construction algorithm for dictionary learning based on PCA or K-Means clustering to model the background and uses low rank and sparsity for constraints to derive a representation coefficient matrix. This approach allows for the exploration of information pertaining to both the overall context and local details, resulting in enhanced discrimination between the non-targets and anomalies \cite{Heesung2005TGRS, Zhou2016TGRS}.

It is not difficult to find that the performance of these methods for background estimation above is strongly associated with the constructed dictionary coefficients. However, the dictionaries constructed using conventional methods exhibit constrained expressive capability in accurately characterizing the real background, particularly in scenarios involving complex backgrounds \cite{8365806,gao2023multi}. Specifically, such methods express the overall context by solving for the lowest low-ranking degree of the dictionary coefficients, while the dictionary atoms have fixed initial values and usually remain constant \cite{hu2022hyperspectral}. However, the method limits the range of effective search space of dictionary coefficients, resulting in a certain upper limit on the expression performance of the algorithm, and a larger search space of coefficients can be obtained for more effective solutions only when dictionary atoms are changed. Taking these factors into account, it is imperative to construct a comprehensive and efficient dictionary.

\begin{figure*}[!t]
	  \centering
			\includegraphics[width=0.9 \textwidth]{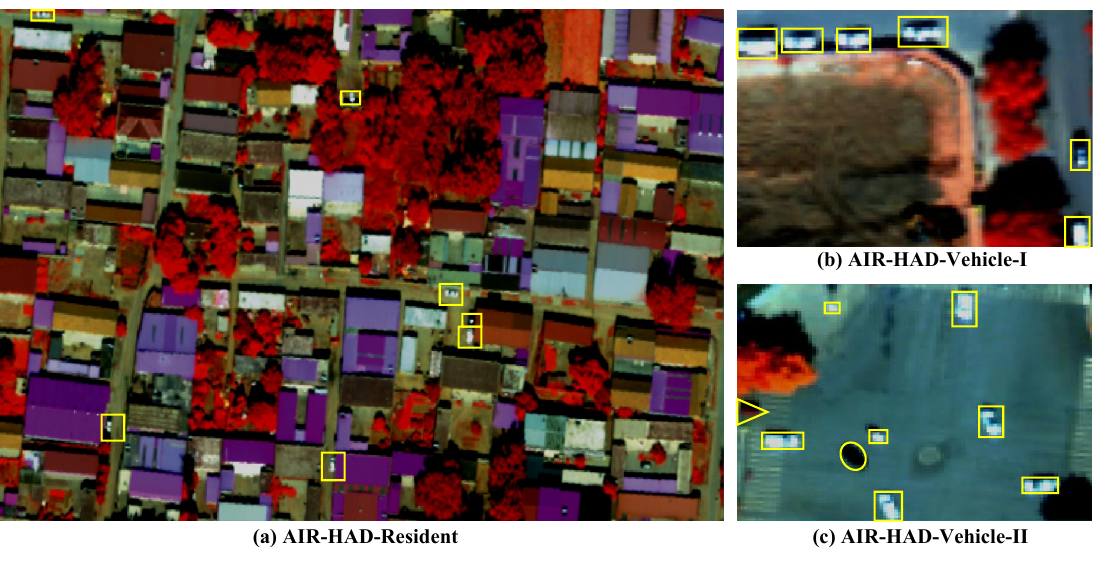}
        \caption{Visualization of AIR-HAD data set, and anomalous targets are shown marked by yellow boxes. (a)-(c) denote the data sets AIR-HAD-Resident, AIR-HAD-Vehicle-I, and AIR-HAD-Vehicle-II, respectively.}
\label{fig2}
\end{figure*}

While data-driven deep learning based techniques, including Convolutional Neural Networks (CNN) \cite{yang2019hyperspectral, wu2023uiu, hong2023cross}, and Autoencoders (AE) \cite{2020Exploiting, chang2020hyperspectral, 2021Self, wang2022auto}, have initially improved the intelligence of unsupervised HAD, they still lack sufficient interpretability. These approaches heavily depend on network architectures as their primary modules, which restricts their interpretability. However, in recent years, there has been increasing interest in deep unfolding or unrolling techniques as a potential solution to design interpretable neural networks \cite{9829013, 2017arXiv170508041D}. Initially, these techniques were utilized for mastering $\mathit{l}_1$ sparse coding \cite{2010Learning}.  In unfolding networks, the fundamental architecture is based on the solution process of conventional physically meaningful models. A series of manually tuned parameters are transformed into trainable parameters for the network \cite{9055228, 9444347}. Specifically, deep unfolding techniques involve mapping each step of the model optimization and solving process in HAD tasks to the network, converting each iteration into a series of network operations, and subsequently combining this series of network operations. It is evident that deep unfolding maximizes the advantages of both data-driven and model-driven techniques \cite{8953470}. More importantly, \cite{8550778, 8578294} have demonstrated that an unfolding network derived from an iterative algorithm can outperform the iterative algorithms for traditional models. This motivates further investigation into whether the HAD performance can be enhanced by combining the strengths of physical-model prior and data-driven techniques.

Given the aforementioned challenges, the main focus of this study is to develop an extensive and challenging dataset to enable thorough evaluation and benchmarking of Hyperspectral Anomaly Detection (HAD) algorithms. To achieve this objective, we procure diverse hyperspectral datasets of varying sizes, encompassing different target types and complexity levels, along with accurately annotated anomaly target ground truth maps. With the support of these data sets, we propose a deep unfolding LRR network with a learnable dictionary (LRR-Net$^+$) that guides the learning process of deep networks by using the physical model as a prior and gives a theoretical basis for ``black box'' operations. The approach establishes an explicit mapping relationship between the physical model and the deep network, resulting in the development of an interpretable work. The present study makes noteworthy contributions in the following aspects:

\begin{itemize}     
     \item  We introduce a novel subspace learning approach that employs simultaneous alternating updates of dictionary atoms and coefficients. By dynamically adjusting the search range of the optimal solution space for coefficients, we enhance the background estimation capability. This is achieved without fixing the upper limit of limiting coefficients, setting it apart from existing methods.
     \item  We show that the LRR physics-model-based unfolding architecture combined with a subspace learning block, named LRR-Net$^+$, can autonomously learn the coefficients of the physical model and provide a deep prior with interpretability. Furthermore, we have tailored an ADMM-based optimization framework to efficiently and accurately solve the optimization solution problem. 
     \item  We build a high-quality and large-scale benchmark dataset for testing the robustness of the HAD task in complex scenarios, AIR-HAD for short. These datasets have been publicly published.      
\end{itemize}

\begin{table*}[!t]
\centering
	\caption{ Selected parameters of the PUBLIC Data SETS in open-source datasets in HAD.}
    \scalebox{1.0}{
	\begin{tabular}{c|c|c|c|c|c}
		\toprule
		\multirow{1}{*}   & Data Size &Spatial Resolution & Anomalous Description& Anomalous & Anomalous Ratio\\
          &  (pixels) & (m) &  & Pixels & (\%) \\
\hline \hline
                  AIR-HAD-Resident & 300 $\times$ 500 & 0.525 & Private car & 142 & 0.09  \\ 
           \hline 
                  AIR-HAD-Vehicle-I & 80 $\times$ 120 & 0.525 & Private car, Truck & 166 & 1.73  \\
           \hline 
                 AIR-HAD-Vehicle-II & 80 $\times$ 100 & 0.525 & Private car & 124 & 1.55  \\ 
		\bottomrule[1.5pt]
	\end{tabular}
 }
    \label{Tab2}
\end{table*}

The subsequent sections of this paper are structured as follows: Section \ref{Method} introduces the establishment of the problem model and its optimization method, and describes the expansion method of the proposed network. Subsequently, the obtained results are presented and analyzed in Section \ref{Experiments}. Finally, the conclusions and potential future research directions are discussed in \ref{Conclusion}.

\section{The AIR-HAD Dataset} \label{Dataset}
With the help of datasets that prove challenging and excellent, the development of the discipline can be accelerated. To facilitate technical breakthroughs and enhance the generalizability of HAD applications, we have undertaken the production of high-quality, large-scale HAD benchmark datasets.

\subsection{Data Collection}
The AIR-HAD dataset was captured on September 22, 2018, using a high-definition specialized aerial system developed by the Shanghai Institute of Optics and Fine Mechanics, Chinese Academy of Sciences. The system employed a full-spectrum multimodal imaging spectrometer that utilizes visible and near-infrared spectrometers. The aerial imagery was acquired from a flight height of 2000 meters. The payload exhibited a total field of view measuring 40.6°, which included an instantaneous field of view of 0.25 mrad, along with an effective scanning pixel count of 2834.

The UAV-borne hyperspectral data underwent preprocessing, which involved employing the ENVI software provided by the instrument manufacturer to perform radiometric calibration and geometric correction. During radiometric calibration, the raw digital numbers (DN) were converted to radiance values using the sensor's calibration parameters. Subsequently, a geometric correction was carried out using the collinearity equation, along with the position and attitude information recorded by the GPS/IMU (Global Positioning System/Inertial Measurement Unit) module.

The complete image of this dataset consists of a grid with dimensions of $3751 \times 2746$ pixels, encompassing 253 spectral bands ranging from 400 to 1000 nm. It primarily presents information about the residential areas located within the Xiong'an area in Beijing, China. 

\subsection{Labeling of Abnormal Information}
Our datasets primarily include the three scenes: urban residential areas, intersections, and building streets. In these scenes, multiple private cars are considered anomalous targets. In the upcoming sections, we will provide a detailed explanation of the specific anomaly annotation process. Within these annotations, the anomalous targets are highlighted within the yellow boxes in Figure 2, and different shapes represent varying sizes and models of cars.

\begin{itemize}     
     \item  Aerospace Information Research (AIR)--Residential dataset (AIR-HAD-Resident). The first dataset was collected over an urban residential area, which contains 300 $\times$ 500 pixels and 253 spectral bands in the spectral range of 400-1000 nm. This dataset is also much larger than the commonly used publicly available HAD dataset. The number of anomalous pixels is 142, which represents 0.094\% of the entire image.
     \item Aerospace Information Research (AIR)--Vehicle dataset (AIR-HAD-Vehicle). The second dataset contains two sub-images, namely AIR-HAD-Vehicle-I and AIR-HAD-Vehicle-II, depicting scenes of a certain intersection and architectural street, where various vehicle models such as private cars and forklifts are considered anomalies. AIR-HAD-Vehicle-I has a size of $80 \times 120$ pixels, with 166 anomalous pixels, accounting for 1.73\% of the image. AIR-HAD-Vehicle-II has a size of $80 \times 100$ pixels, with 124 anomalous pixels, representing 1.55\% of the entire image.   
\end{itemize}

Regarding the statistical analysis and annotation of anomaly information, our work is based on the criteria proposed in \cite{chang2016}: 1) No prior knowledge of anomaly existence. Since there is no prior knowledge about the anomalies, this criterion is satisfied. 2) Low occurrence probability and 3) Insignificance in spectral statistics. The assessment of these two criteria involves computing the ratio of anomalous pixels to the overall pixel count across the entire image. The proportion of anomalous pixels within the AIR-HAD-Resident dataset is 0.09\% as indicated in Tab. \ref{Tab2}, and for the AIR-HAD-Vehicle dataset, it is 1.73\% and 1.55\% respectively. The proportion of anomalies in each dataset is very low, thus satisfying criteria 2 and 3. As for the final criterion, 4) Small population size, the pixel thresholds for the AIR-HAD-Resident and AIR-HAD-Vehicle datasets are 150,000, 9,600, and 8,000 respectively, which also meet this criterion.


\section{LRR-Net$^+$: LRR Meets Deep Unfolding} \label{Method}

\begin{figure*}[!t]
	  \centering
			\includegraphics[width=0.9\textwidth]{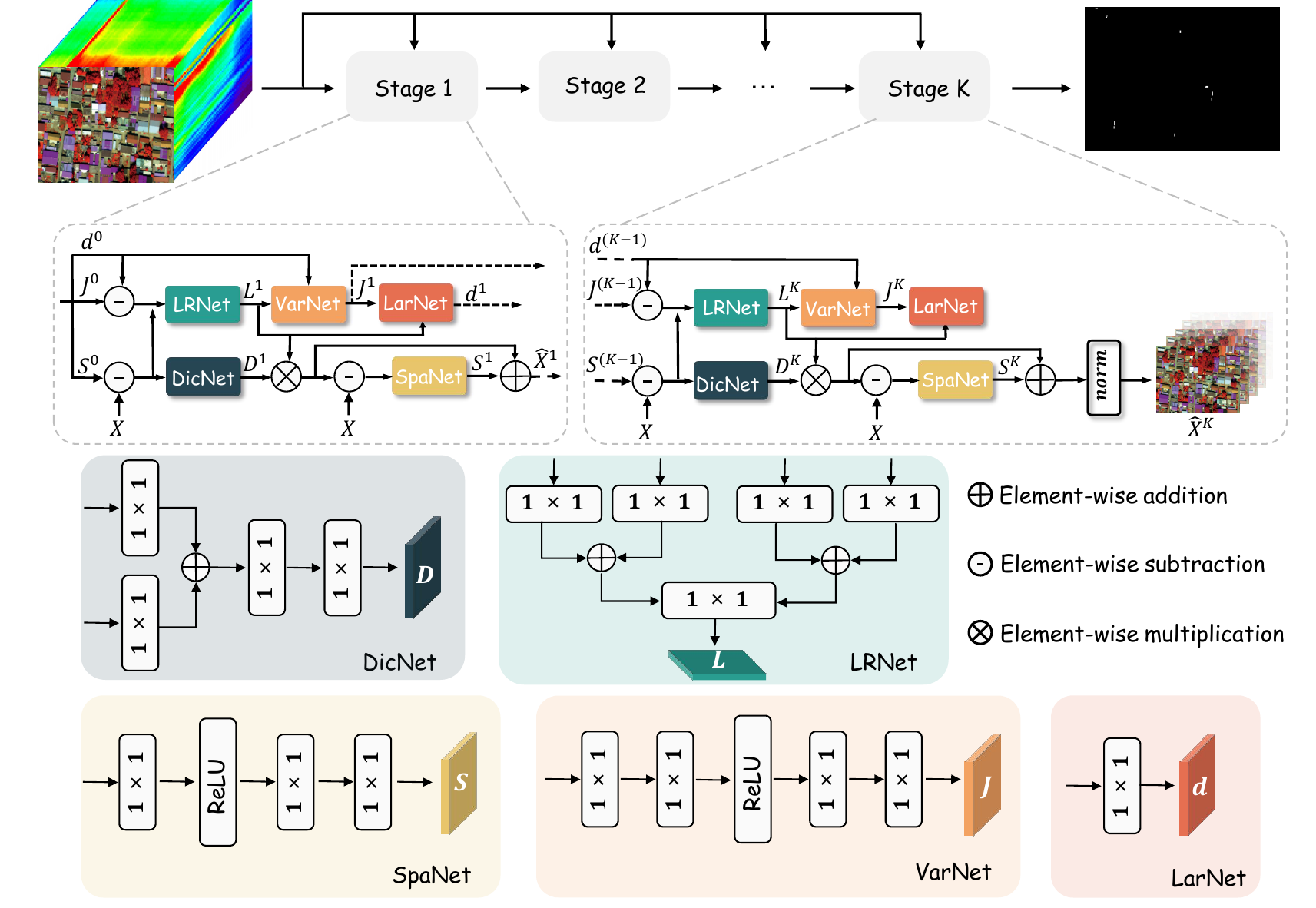}
        \caption{An illustration of the iterative learning process of the deep unfolding LRR model in the proposed LRR-Net$^+$ network. The to-be-estimated variables $\mathbf{D}$, $\mathbf{L}$, $\mathbf{S}$, $\mathbf{J}$, and $\mathbf{d}$ denote the dictionary atoms, dictionary coefficients, anomaly vectors, auxiliary variables, and Lagrangian multipliers, respectively.}
\label{fig3}
\end{figure*}

\subsection{Method Overview}
To improve the efficacy of the dictionary learning for the background estimation and break the search limitation caused by fixed values, we expand the search range of the optimal solution by alternatively updating the dictionary atoms and coefficients simultaneously. Using the alternating direction multiplication method (ADMM) optimization framework \cite{Boyd2011FTML, hong2019augmented}, there is sufficient accuracy and fast convergence speed in the solution process. To this end, we design an interpretable HAD network, named LRR-Net$^+$, that offers a comprehensive and explicit end-to-end process by establishing a direct mapping between regularization parameters and trainable deep neural network parameters, achieved through a meticulous step-by-step association of conventional modules and operators. The use of deep unfolding techniques, coupled with physical prior knowledge, has excellent potential for unsupervised learning. The flowchart of the proposed LRR-Net$^+$ architecture is illustrated in Fig. \ref{fig3}.

\subsection{Model Formulation and Optimization}
 According to \cite{Liu2013TPAMI}, there usually exists a strong correlation among the background pixels, i.e., they satisfy low rankness, while the anomalous target has the property of sparsity. Based on this, Let $\mathbf{X}=[\mathbf{x}_1,...,\mathbf{x}_i,...,\mathbf{x}_M]\in\mathbb{R}^{M \times B}$ represent the measured HSI matrix with $\mathit{M}=\mathit{C}\times\mathit{H}$ pixels, where $\mathbf{x}_i$ $(i=1,..., M)$ denotes the $i$-th spectral vector with $B$ dimensions. In this context, we assume that the given matrix can be expressed as a combination of a background component $\mathbf{D}\mathbf{L}$ and a target component $\mathbf{S}$, where $\mathbf{D}$ represents the dictionary containing the background samples, and $\mathbf{L}$ denotes the representation coefficients for the background. The target component $\mathbf{S}$ corresponds to the anomalies or deviations from the background. This assumption allows us to model the HAD problem and facilitate the separation of anomalies from the background in the given hyperspectral data.
 
\begin{equation}
\label{eq1}
\begin{aligned}
     \mathbf{X}=\mathbf{D}\mathbf{L}+\mathbf{S}.
\end{aligned}
\end{equation}
Usually, the estimation performance of the image background is highly correlated with the construction of the dictionary, which determines the search range of the solution space.

As a common practice, we replace the rank function with the nuclear norm, while we improve the representation performance of the algorithm by changing the atoms and coefficients of the dictionary. Additionally, Drawing inspiration from the unmixed model, we adopted a less restrictive approach to describe the optimization model for Eq. (\ref{eq2}). Consequently, we formulated the anomaly problem as follows

\begin{equation}
\label{eq2}
\begin{aligned}
\underset{\mathbf{D}, \mathbf{L}, \mathbf{S}}{\textrm{min}}\; & \frac{1}{2} || \mathbf{X} - \mathbf{D} \mathbf{L} -\mathbf{S}  ||_{\F}^2 + \frac{\lambda_1 }{2}\norm{\mathbf{D}}_{\F}^2 + \lambda_2 \norm{\mathbf{L}}_{*} + \lambda_3 \norm{\mathbf{S}}_{2,1},
\end{aligned}
\end{equation}
where $\lVert\cdot\rVert_{\F}$ represents Frobenius norm, and in the context of the nuclear norm. Let $\|\mathbf{S}\|_{2,1} = \sum_{i=1}^N \sqrt{\sum_{j=1}^M (\mathit{s}_{i,j})^{2}}$ denote the $\mathit{l}_{2,1}$-norm of the anomaly matrix $\mathbf{S}$, where $\mathbf{S}_i$ represents the $i$th column of the anomaly matrix $\mathbf{S}$. Moreover, $\lambda_1 \geq 0$, $\lambda_2 \geq 0$, and $\lambda_3 \geq 0$ represent the penalty parameters that control the relative importance of different terms in the model, which could be chosen according to properties of the different norms, or tuned empirically.

The optimization problem is generally non-convex, making it challenging to find the global minimum directly in one step due to the difficulty of obtaining an analytical solution. However, our approach seeks to obtain local optimal solutions by iteratively optimizing separable convex subproblems involving the variables $\mathbf{D}$, $\mathbf{L}$, and $\mathbf{S}$ to be estimated. To optimize the problem, an ADMM-based solver is designed, introducing an auxiliary variable $\mathbf{J}$ to replace $\mathbf{L}$ in the constraint term of Eq.(\ref{eq2}), leading to $\mathbf{L}=\mathbf{J}$ and making the objective function separable.

The Eq. (\ref{eq2}) is further reformulated into its corresponding augmented Lagrangian function
\begin{equation}
\label{eq3}
\begin{aligned}
      \hspace{-0.4cm}\mathscr{L}_{U}&\left(\mathbf{D},\mathbf{L},\mathbf{S}, \mathbf{J}, \mathbf{d} \right) =\frac{1}{2}\norm{\mathbf{X} - \mathbf{D} \mathbf{L} -\mathbf{S}}_{\F}^2  \\
     &  + \frac{\lambda_1}{2} \norm{\mathbf{D}}_{\F}^2 + \lambda_2 \norm{\mathbf{J}}_{*} + \lambda_3 \norm{\mathbf{S}}_{2,1} + \frac{\mu}{2}\norm{\mathbf{L} - \mathbf{J} + \mathbf{d}}_{\F}^2,
\end{aligned}
\end{equation} 
where the variable $\mathbf{d} \geq 0$ denotes the Lagrange multiplier and $\mathbf{\mu} \geq 0$ is the  regularization parameter.

Within the ADMM optimization framework, we can effectively address problem (\ref{eq3}) by iteratively minimizing the objective function $\mathscr{L}$ with respect to the variables $\mathbf{D}$, $\mathbf{L}$, $\mathbf{S}$, $\mathbf{J}$, and $\mathbf{d}$, while keeping the other variables fixed.

\begin{itemize}   
     \item  \textit{Optimization concerning} $\mathbf{D}$: The optimization pertaining to variable $\mathbf{D}$ can be formulated as a least-square regression problem with a regularizing term following the Tikhonov-Phillips regularization. Therefore, this subproblem can be expressed as follows:
     
\begin{equation}
\label{eq4}
\begin{aligned}
      \underset{\mathbf{D}}{\textrm{min}}\;\frac{\lambda_1}{2} \norm{\mathbf{D}}_{\F}^2 + \frac{1}{2}|| \mathbf{X} - \mathbf{D}\mathbf{L} - \mathbf{S} ||^2_{\F}.
\end{aligned}
\end{equation} 

Therefore, the closed-form solution for this problem is as follows
\begin{equation}
\label{eq5}
\begin{aligned}
      \mathbf{D} \leftarrow  
 (\mathbf{L}^{\T}\mathbf{L}+{\lambda_1}\mathbf{I})^{-1}\mathbf{L}^{\T}(\mathbf{X}-\mathbf{S}),
\end{aligned}
\end{equation} 
where $\mathbf{I}$ is a unit matrix, and all the following occurrences are similar.

 \item  \textit{Optimization with respect to} $\mathbf{L}$: The optimization problem of the variable $\mathbf{L}$ can be formulated as follows
\begin{equation}
\label{eq6}
\begin{aligned}
      \underset{\mathbf{L}}{\textrm{min}}\;\frac{1}{2}|| \mathbf{X} - \mathbf{D}\mathbf{L} - \mathbf{S} ||^2_{\F} + \frac{\mu}{2}\norm{\mathbf{L} - \mathbf{J} + \mathbf{d}}_{\F}^2.
\end{aligned}
\end{equation} 

the analytical solution for this can be expressed as follows
\begin{equation}
\label{eq7}
\begin{aligned}
      \mathbf{L} \leftarrow  
 (\mathbf{D}^{\T}\mathbf{D}+{\mu}\mathbf{I})^{-1}[\mathbf{D}^{\T}(\mathbf{X}-\mathbf{S}) + \mu(\mathbf{J} - \mathbf{d})],
\end{aligned}
\end{equation} 
        
     \item  \textit{Optimization with respect to} $\mathbf{S}$: The optimization task related to variable $\mathbf{S}$ can be formulated as follows
\begin{equation}
\label{eq8}
\begin{aligned}
      \underset{\mathbf{S}}{\textrm{min}}\;\mathbf{\lambda_3}|| \mathbf{S}||_{2,1}  + \frac{1}{2}|| \mathbf{X} - \mathbf{D}\mathbf{L} - \mathbf{S}||^2_{\F},
\end{aligned}
\end{equation} 
it has an approximate solutions similar to the $\mathit{l}_{2,1}$-norm .

     \item  \textit{ Optimization with respect to} $\mathbf{J}$:
\begin{equation}
\label{eq9}
\begin{aligned}
      \underset{\mathbf{J}}{\textrm{min}}\;\mathbf{\lambda_2}|| \mathbf{J}||_*  + \frac{\mu}{2}\norm{\mathbf{L} - \mathbf{J} + \mathbf{d}}_{\F}^2.
\end{aligned}
\end{equation}  
The solution to this problem can be obtained using the Singular Value Thresholding (SVT) operator \cite{Liu2013TPAMI}
 
  1) Singular value decomposition (SVD) is performed by
\begin{equation}
\label{eq10}
\begin{aligned}
      [\mathbf{U}, \mathbf{\sum}, \mathbf{V}] = svd((\mathbf{L} - \mathbf{d})-{\mathbf{\lambda_2}/\mu}),
\end{aligned}
\end{equation}
such that $(\mathbf{L} - \mathbf{\mathbf{d}})-{\mathbf{\lambda_2}/\mu} = \mathbf{U}\mathbf{\sum}\mathbf{V}^{\T}$. 

  2) For each ${\lambda_2 /\mu} \geq 0$
\begin{equation}
\label{eq11}
\begin{aligned}
      \mathbf{J} \leftarrow \mathbf{U} \times \mathbf{\sum} \times \mathbf{V}^{\top},
\end{aligned}
\end{equation}
where $\{\cdot\}_+=max(0,x)$, and $\sum=diag(\{\mathbf{L} - \mathbf{d}\}_+)$.

Eq.(\ref{eq9}) can be approximated and computed.

     \item  \textit{ Updating with respect to Lagrange multiplier} $\mathbf{d}$:
\begin{equation}
\label{eq12}
\begin{aligned}
      \mathbf{d} \leftarrow \mathbf{d}- \mu (\mathbf{L}  - \mathbf{J}),
\end{aligned}
\end{equation} 
where regularization parameter $\mu$ is update by
\begin{equation}
\label{eq13}
\begin{aligned}
      \mu \leftarrow \mathbf{min (\rho\mu, \mu_{\max})},
\end{aligned}
\end{equation} 
in this case, $\rho > 1$ and $\mathbf{\mu}_{\max}$ are used to represent the scaling factor and the maximum value of $\mathbf{\mu}$ during a specific number of iterations, respectively.  
\end{itemize}

\begin{algorithm}[!t]
\caption{Optimization Process for LRR Model in LRR-Net$^{+}$}
\KwIn{Observed HSI $\mathbf{X}$, $\mathbf{D}$, $\mathbf{L}_{0}$, $\mathbf{S}_{0}$, and parameters $\rho$.}
\KwOut{Background $\mathbf{L}$, Anomaly map $\mathbf{S}$.}
\textbf{Initialization}:
$\mathbf{J}=\mathbf{0}$, $\mathbf{d}=\mathbf{0}$, $\mathbf{\Lambda} =\mathbf{0}$, $\mathbf{\Gamma} =\mathbf{0}$, $\mathbf{\Theta}=\mathbf{0}$, $\lambda_1=0.5$, $\lambda_2=1.2\times10^{-5}$, $\lambda_3=10^{-5}$, $\mu=1$\\
\While{$0 \leq k \leq K$ or meet the convergence conditions}
 {
          Fix other variables and update $\mathbf{D}$ by Eq. (\ref{eq14})\\
          Fix other variables and update $\mathbf{L}$ by Eq. (\ref{eq15})\\
          Fix other variables and update $\mathbf{S}$ by Eq. (\ref{eq16}) \\
          Fix other variables and update $\mathbf{J}$ by Eq. (\ref{eq19})\\
          Fix other variables and update $\mathbf{d}$ by Eq. (\ref{eq21})\\
         Check the convergence conditions:\\
         \eIf{$\norm {\mathbf{L}-\mathbf{J}}_{\F}<\varepsilon$, $||x^{k}-\hat{x}^{k-1}||_F^2<\zeta$}
         {
           Stop iteration;
         }
         {
         $k\leftarrow k+1$;
         }
 }
\end{algorithm}

\subsection{Deep Unfolding Solver into Neural Network}
To develop an interpretable deep network architecture for the HAD task, we employ a multi-stage unfolding approach, which involves iteratively expanding the network by replicating the iterative steps. This allows us to construct a comprehensible and interpretable deep network. In particular, the neural network can be conceptualized as comprising five sub-networks, each representing a solution step for one variable of the ADMM algorithm. Together, these sub-networks form one complete cycle of the neural network. The network consists of $K$ stages, each corresponding to one iteration in the iterative algorithm used to solve Eq. (\ref{eq2}). This architecture is illustrated in Fig. \ref{fig3}.

\textbf{ 1) {Dictionary Update (DicNet)}}. According to Eq. (\ref{eq5}), we introduce a learnable parameter $\mathbf{\lambda}_1$ to derive the updated iterative step, which can be expressed as follows
\begin{equation}
\label{eq14}
\begin{aligned}
       \mathbf{D}=\mathbf{DicNet}_{\mathbf{\Lambda}}(\mathbf{X} - \mathbf{S}),
\end{aligned}
\end{equation}
where $\mathbf{\Lambda}=(\mathbf{L}^{\T}\mathbf{L}+2\lambda_1\mathbf{I})^{-1}\mathbf{L}^{\T}$. Note that the work update of $\mathbf{DicNet}_{\mathbf{\Lambda}}$ is obtained by multiple linear layers and the ``Add'' operation.

\textbf{ 2) {Low-Rank Update (LRNet)}}. This approach is formulated by unfolding $\mathbf{L}$, which enables the representation of the $k$th iterate $\mathbf{L}^{k}$ using the $k$th estimates of $\mathbf{D}^k$, $\mathbf{S}^{(K-1)}$, $\mathbf{d}^{(K-1)}$, $\mathbf{J}^{(K-1)}$, and $\mathbf{X}$ in the following manner
\begin{equation}
\label{eq15}
\begin{aligned}
     \mathbf{L}=\mathbf{LRNet}_{\mathbf{\Gamma}, \mathbf{\Theta}}(\mathbf{X} - \mathbf{S}, \mathbf{J} - \mathbf{d}),
\end{aligned}
\end{equation}
where $\mathbf{\Gamma}=(\mathbf{D}^{\T}\mathbf{D}+\mu\mathbf{I})^{-1}\mathbf{D}^{\T}$, $\mathbf{\Theta}=(\mathbf{D}^{\T}\mathbf{D}+\mu\mathbf{I})^{-1}\mu$. $\mathbf{LRNet}_{\mathbf{\Gamma}, \mathbf{\Theta}}$ is formed by integrating multiple linear layers in a similar manner as $\mathbf{DictNet}_{\mathbf{\Lambda}}$, with $\mathbf{\Gamma}$ and $\mathbf{\Theta}$ being two trainable parameters utilized to derive the updated iterative step. The new iterative step is obtained by utilizing the two learnable parameters $\mathbf{\lambda}_1$ and $\mathbf{\lambda}_2$.

\textbf{ 3) {Sparse Update (SpaNet)}}. To approximate the soft-threshold operator, we utilize the linear rectification function (ReLU) to enforce the non-negativity constraint on the output sparsity $\mathbf{S}$. Consequently, we can reformulate Equation (\ref{eq8}) as follows

\begin{equation}
\label{eq16}
\begin{aligned}  
\mathbf{S}^{(k)}=\mathbf{SpaNet}_{\lambda_3}(\mathbf{X} - \mathbf{D}\mathbf{L}),
\end{aligned}
\end{equation}  
where $\mathbf{\lambda}_3$ as a learnable parameter, and $\mathbf{SpaNet}_{\lambda_3}(\mathbf{\cdot})$ represents the soft-thresholding operator, which can be expressed as
\begin{equation}
\label{eq17}
\begin{aligned}
\mathbf{sign}(\lVert\rm{s}\lVert_{2})\mathbf{max}(0, \lVert\rm{s}\lVert_{2}-\mathbf{\lambda_3)},
\end{aligned}
\end{equation}  
where $\lVert\cdot\rVert_{2}$ represents the $\mathit{l}_{2}$-norm computation for each column.

Due to the similarity in functionality between the soft-thresholding operator and the ReLU function, the update for Eq. (\ref{eq8}) in LRR-Net$^+$ can be accomplished by
\begin{align}\label{eq18}
\frac{\mathbf{X}-\mathbf{D}\mathbf{L}}{\lVert\mathbf{X}-\mathbf{D}\mathbf{L}\rVert_2}\rm{ReLU}(\lVert\mathbf{X}-\mathbf{D}\mathbf{L}\rVert_2-\lambda_3).
\end{align}

\textbf{ 4) {Auxiliary Variable Update (VarNet)}}. Likewise, the calculation of intermediate variables $\mathbf{L}^{(K-1)}$ and $\mathbf{d}^{(K-1)}$ is required in the current iteration, and it can be expressed as follows
\begin{equation}
\label{eq19}
\begin{aligned}
    \mathbf{J}^{(k)}= \mathbf{VarNet}_{\mathbf{\theta}}(\mathbf{L} - \mathbf{d}),
\end{aligned}
\end{equation}
where $\mathbf{VarNet}_{\mathbf{\theta}}{(\mathit{\cdot})}$ is solved via applying the proximal operator of the nuclear norm, and we write ${\mathbf{\lambda_2}/\mu}$=$\mathbf{\theta}$.

It is worth noting that the $\rm{ReLU}$ operation is a component-wise rectified linear unit, which aligns with the proximal operator of the nuclear norm as described in \cite{Goodfellow-et-al-2016}. As a result, Eq. (\ref{eq19}) can be expressed as follows:
\begin{equation}
\label{eq20}
\begin{aligned}
    \mathbf{U} \times \mathbf{ReLU} \{ \textbf{diag} (\mathbf{L} - \mathbf{\mathbf{d}}) - \theta\mathbf{I} \} \times \mathbf{V}^{\T}.
\end{aligned}
\end{equation}

\textbf{ 5) {Lagrange Multiplier Update (LarNet)}}. To update the Lagrange multiplier variable $\mathbf{d}$ in each iteration, we introduce a learnable parameter $\mathbf{\mu}$ in Eq. (\ref{eq12}), resulting in a revised iterative form 
\begin{equation}
\label{eq21}
\begin{aligned}
       \mathbf{d}^{(k)}=\mathbf{LarNet}_{\mu}(\mathbf{L}-\mathbf{J}),
\end{aligned}
\end{equation}
where, we compute $\mathit{d}_{\mu}$ as $\mathbf{d} - \mu (\mathbf{L} - \mathbf{d})$, where $\mathbf{d}$ serves as a learnable parameter, providing added flexibility to the process.

Finally, the estimated dictionary coefficient $\hat{\mathbf{L}}$ = $\mathbf{J}^{K}$ of the final output is multiplied by the atom $\mathbf{D}^K$, and the abnormal pixels $\mathbf{S}^K$ to obtain the reconstructed HSI $\hat{\mathbf{X}}$.

In addition to the introduction of the network framework mentioned above, it is worth noting that our empirical findings have shown that performing normalization after each iteration significantly benefits the optimization and convergence of the network. This advantage has a dual effect. Particularly, it ensures stable model training, thereby mitigating concerns related to gradient vanishing and explosion that might arise during the network training phase. Second, it accelerates the convergence speed of network learning, leading to the discovery of better solutions.

\subsection{Initialization and Training}
Conventionally, neural networks undergo training by employing a process of random initialization for their parameters. However, to improve the speed and accuracy of the training process, LRR-Net$^+$ incorporates a physical model as prior guidance. We utilize an approach for parameter initialization in the neural network, which involves using the original parameters linked to the ADMM algorithm, as illustrated below.

1) The initial values of the variables $\mathbf{D}$ and $\mathbf{L}$ are calculated by K-means clustering, correlated with DicNet and SpaNet, and involved in subsequent iterative updates.

2) The initialization and training of the learnable parameters $\mathbf{\Lambda}$ and $\mathbf{\Gamma}$, $\mathbf{\Theta}$ are associated with the update of the variables $\mathbf{D}$ and $\mathbf{L}$, which are related to parameters $\mathbf{\lambda_1}$ and $\mu$ in the ADMM algorithm, respectively. 

3) We initiate the parameter $\lambda_3$ for the S-update component in each layer, and its initialization is intricately connected to the parameter from the preceding step, effectively regulating the influence of anomalous background.

4) The initialization of the parameter $\theta$, which governs the VarNet-update component, involves the use of ${\mathbf{\lambda_2}/\mu}$, where $\lambda_2$ and $\mu$ are linked to parameters from the ADMM algorithm.

5) Lastly, we update the parameter $\mu$ using the expression $min (\rho\mu, \mu_{\max})$. This adjustable parameter serves as a scaling factor, allowing us to adapt and optimize it according to the particular characteristics of the HAD task.

Throughout the entire update iteration process of the deep unfolded network, all algorithm parameters remain capable of being adjusted.

Considering the lack of labeled data in the HAD task domain, we address this issue by employing an unsupervised training approach. The loss function used in this work is based on the standard mean-squared error (MSE), which measures the discrepancy between the predicted data $\hat{\mathbf{X}}$ and the original input DATA $\mathbf{X}$. This optimization process aims to enforce the estimated anomaly values to be closer to the actual anomaly pixels $\mathbf{S}$. The loss function is given by

\begin{equation}
\label{eq22}
\begin{aligned}
\text{R}_{loss} = \mathbf{MSE(\mathbf{X}, \hat{\mathbf{X}})}=\frac{1}{N}\sum_{i=1}^N\lVert\mathbf{x}_i-\hat{\mathbf{x}}_i\rVert_2^2,
\end{aligned}
\end{equation}
where $\mathbf{X}$ consists of N reflectance spectra $\mathbf{x}_{i}$, $\hat{\mathbf{x}_i}$ corresponds to the estimated value of $\hat{\mathbf{X}}$, which can also be considered as the reconstructed values.

here, $\mathbf{X}$ comprises N reflectance spectra $\mathbf{x}_{i}$, and $\hat{\mathbf{x}_i}$ represents the estimated value of $\hat{\mathbf{X}}$, which can also be interpreted as the reconstructed values.

\begin{table*}[!t]
\centering
	\caption{The AUC scores of ROC curves $(\%)$ with nine methods in different datasets.}
    \scalebox{1.0}{
	\begin{tabular}{c||ccccccc|c}
		\toprule
		\multirow{1}{*} Scenes & Global-RX & Local-RX & CRD & LRASR & CNN & Auto-AD & LRR-Net & \textbf{LRR-Net$^{+}$} \\
		\hline\hline       
        Airport-1 & 82.21 & 95.76 & 90.73 & 79.32 & 89.49 & 90.29 & 94.35 & \bf 95.87 \\  
  
        Airport-2 & 95.26 & 95.37 & 96.09 & 96.89 & 96.97 & 97.32 & 98.67 & \bf 98.72 \\ 
  
        Airport-3 & 92.88 & 95.10 & 89.04 & 88.91 & 92.16 & 96.75 & 98.13 & \bf 98.51 \\
  
        Airport-4 & 94.03 & 92.63 & 89.96 & 87.58 &95.02 & 96.17 & 96.89 & \bf 97.69 \\
          \hline
         Average  & 91.09 & 94.71 & 91.45 & 88.17 & 93.41 & 95,13 & 97.01 & \bf 97.70 \\
         \hline
        AIR-HAD-Vehicle-I & 58.45 & 65.42 & 76.94 & 75.72 & 89.97 & 90.87 & 95.93 & \bf 96.10 \\ 
  
        AIR-HAD-Vehicle-II & 64.14 & 65.10 & 72.76 & 69.29 & 82.16 & 83.24 & 96.13 & \bf 97.56 \\
  
        AIR-HAD-Resident & 69.70 & 68.20 & 62.18 & 58.89 & 72.32 & 79.13 & 80.69 & \bf 83.69 \\
        \hline
         Average & 64.09 & 66.24 & 70.62 & 67.96 & 81.48 & 84.41 & 90.91 & \bf 92.45 \\ 
		\bottomrule[1.5pt]
	\end{tabular}
 }
    \label{Tab4}
\end{table*}

\begin{figure*}[!t]
	  \centering
			\includegraphics[width=0.9\textwidth]{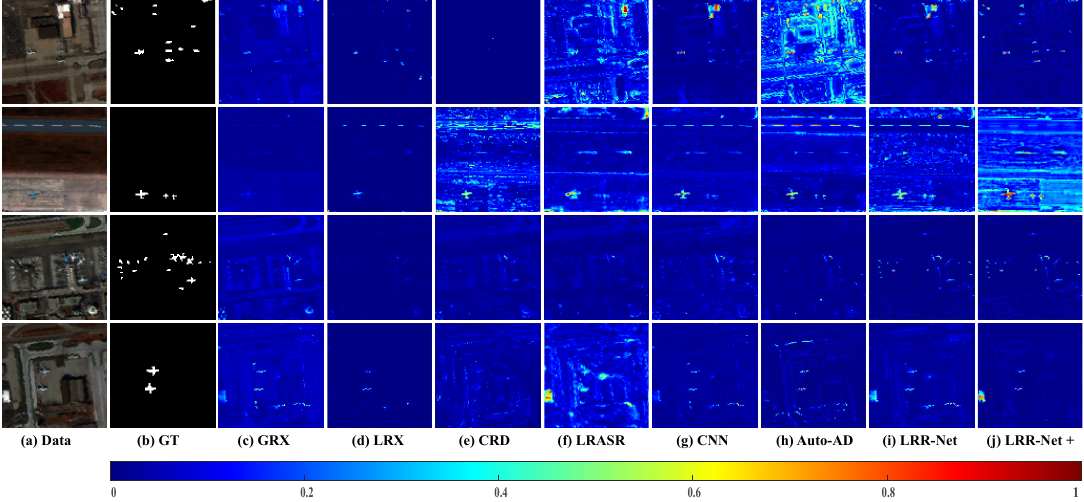}
        \caption{ Image descriptions and detection maps of Airport 1 to 4, showing (a) false-color images, (b) corresponding ground truth, and the detection maps corresponding to (c)–(j) Global-RX, Local-RX, CRD, LRASR, CNN, Auto-AD, LRR-Net, and LRR-Net$^+$ respectively.}
\label{fig4}
\end{figure*}

\section{Experiments} \label{Experiments}
In this section, we commence by presenting the datasets utilized in the experiments, followed by detailing the configuration of the experimental parameters. Then, we compared and analyzed the detection results of several of the best available methods. Finally, we analyze the sensitivity of several network iterations to verify the effectiveness and potential of the deep unfolding technique in the field of HAD.

Unless stated otherwise, all experiments were carried out on a Windows 10 operating system by the PyTorch framework. The experiments were performed using a computational setup consisting of a 3.20-GHz 12th Gen Intel(R) Core(TM) i9-12900KF CPU with 24 cores and an NVIDIA GeForce RTX 3070 GPU.

\subsection{Dataset Description}
\textbf{Airport-Beach-Urban (ABU).} This dataset was captured by the Airborne Visible/Infrared Imaging Spectrometer Sensor (AVIRIS\footnote{ http://aviris.jpl.nasa.gov/}) and the Reflective Optical System Imaging Spectrometer (ROSIS-03) sensor. The ABU dataset, collated and shared by members of Professor Kang's group at Hunan University \cite{7994698}, contains 13 HSIs as airport, beach, and city scenarios, available from its website \footnote{ http://xudongkang.weebly.com/datasets.html}. This paper uses four airport-related datasets obtained in 2010 that cover an area in Gulfport, southern Mississippi, USA. The details of the dataset are given in Tab. \ref{Tab1}.

\textbf{Aerospace Information Research (AIR-HAD).} The dataset compiled and shared in this paper, includes three scenes: residential areas, intersections, and streets, in which multiple vehicles are considered anomalies in urban scenes. Detailed information about the dataset is provided in Table \ref{Tab2}. It is available on the website\footnote{https://sites.google.com/view/danfeng-hong}. The second part of the article provides a detailed description of the statistical analysis and annotation of the anomaly targets.

\begin{figure*}[!t]
	  \centering
			\includegraphics[width=0.9\textwidth]{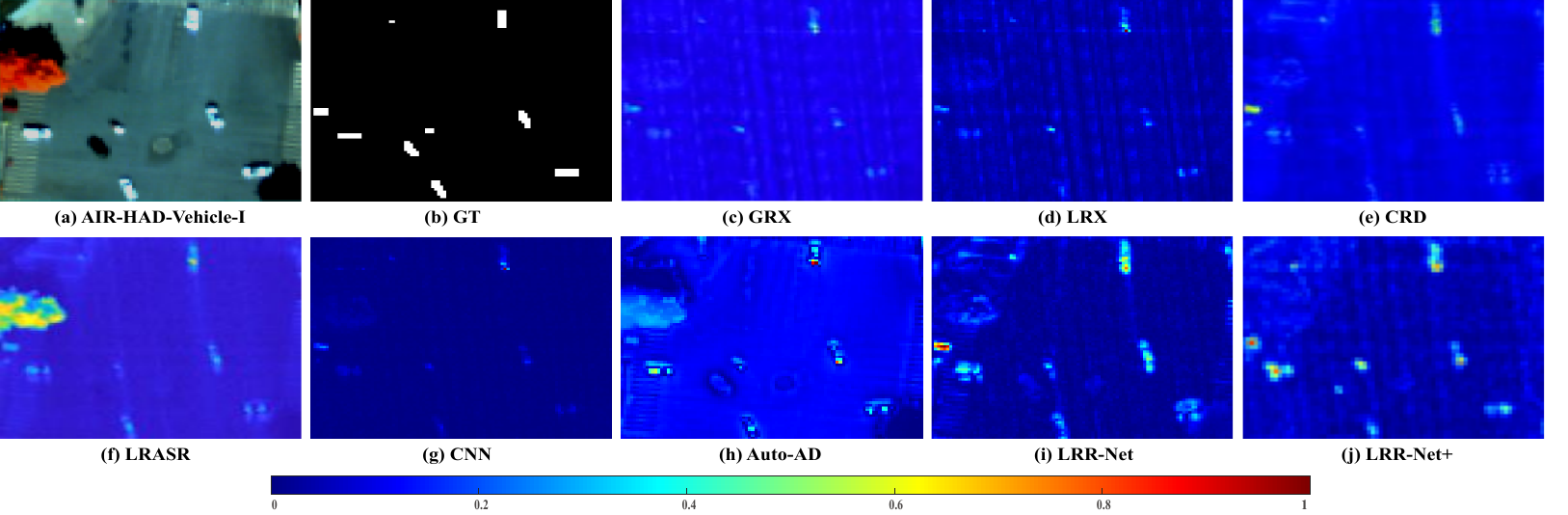}
        \caption{ Image descriptions and detection maps of AIR-HAD-Vehicle-II, showing (a) false-color images, (b) corresponding ground truth, and the detection maps corresponding to (c)–(j) Global-RX, Local-RX, CRD, LRASR, CNN, Auto-AD, LRR-Net, and LRR-Net$^+$ respectively.}
\label{fig6}
\end{figure*}

\begin{figure}[!t]
	  \centering
			\includegraphics[width=0.45\textwidth]{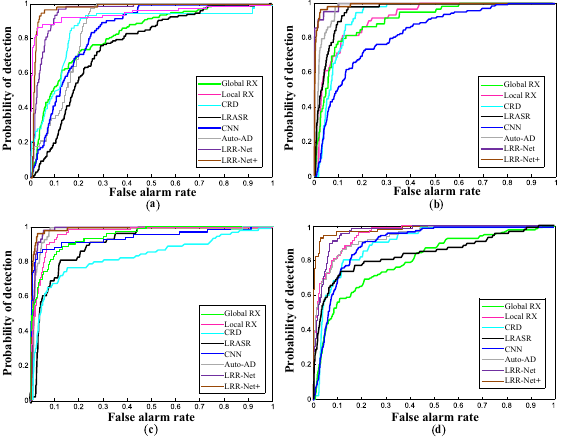}
        \caption{ROC curves obtained by different methods for the Airport 1 to 4 datasets.}
\label{fig5}
\end{figure}

\subsection{Evaluation Metric and Implementation Details}
\begin{figure*}[!t]
	  \centering
			\includegraphics[width=0.9\textwidth]{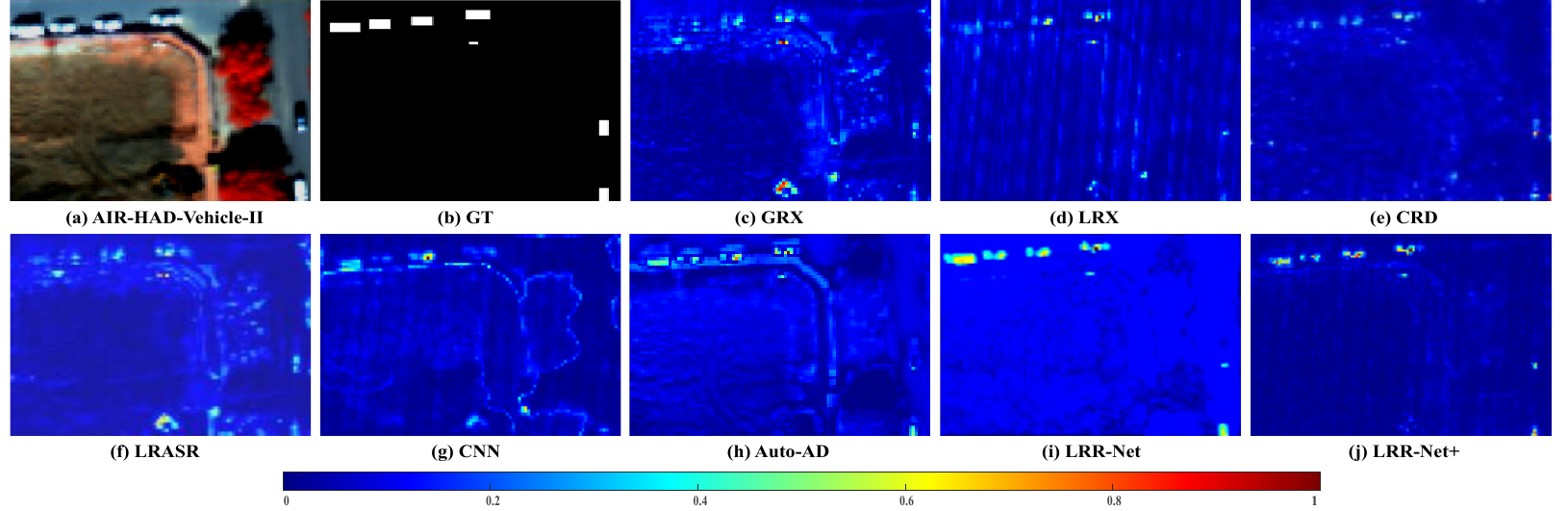}
        \caption{ Image descriptions and detection maps of AIR-HAD-Vehicle-II, showing (a) false-color images, (b) corresponding ground truth, and the detection maps corresponding to (c)–(j) Global-RX, Local-RX, CRD, LRASR, CNN, Auto-AD, LRR-Net, and LRR-Net$^+$ respectively.}
\label{fig7}
\end{figure*}

To assess the detection performance, we employed the Receiver Operating Characteristic (ROC) curve, which illustrates the correlation and trends between the Probability of Detection (PD, i.e., correctly detecting anomalies) and the False Alarm Rate (FAR, i.e., generating false alarms) at different thresholds, as depicted in Fig. \ref{fig5}, and \ref{fig9}. Additionally, we use the Area Under Curve (AUC) to assess the anomaly detection performance of the algorithm quantitatively \cite{Khazai2011_AUC}, as shown in Table \ref{Tab4}. We also present visual color detection maps to provide a more intuitive comparison of the performance of various algorithms, as shown in Fig. \ref{fig4}, \ref{fig6}, \ref{fig7}, and \ref{fig8}. In these maps, the color range from blue to red represents the magnitude of anomaly probability, whereas regions tending towards bright red indicate a higher likelihood of anomalies.

\subsection{Experimental Analysis and comparison}
In this section, based on the provided dataset, we compare our proposed LRR-Net$^+$, with several popular existing HAD algorithms, including G-RX \cite{reed1990adaptive}, L-RX \cite{Molero2013}, CRD \cite{li2014collaborative}, LRSAR \cite{xu2016TGRS_LRSAR}, CNN \cite{CHANG2019TGRS_HSI-DeNet}, Auto-AD \cite{wang2022auto}, and LRR-Net \cite{10136197}, in terms of experiments and performance.

For the specific experimental parameter settings, we referred to the original descriptions of various methods. The G-RX method computes the anomaly targets by computing the squared Mahalanobis distance between the test pixel and the mean value of the entire image background, which serves as a measure for anomaly detection. In L-RX, instead of using the global covariance matrix, the local covariance matrix is computed. For this experiment, we used $\mathit{W}_{in}=[3, 5, 7, 9, 11, 13, 15, 17, 19]$ and $\mathit{W}_{out}=[5, 7, 9, 11, 13, 15, 17, 19, 21, 23]$. The CRD method employs a regularization parameter, $\lambda = 10^{-6}$, for its implementation. The LRASR method is configured with a choice of 15 clusters ($K=15$) and the selection of 20 pixels ($P=20$). Additionally, the regularization parameters $\beta$ and $\lambda$ are fixed at 0.1 and 0.5, respectively. For Auto-AD experiments, we utilized the provided code, employing an encoder-decoder structure with four 1-D max convolutional layers having sizes [C, H, W] = {[64, 1, 9], [128, 1, 9], [256, 1, 9], [512, 1, 9]}. These layers form a symmetric structure with multiple deconvolution layers. The experiment for LRR-Net was conducted using the hyperparameters provided in the reference paper, which were chosen as follows: a learning rate of 0.0015, a batch size of 15, and a total of 10 updates.

\textbf{Airport:} In Fig. \ref{fig4} (a), strong interference information is observed in the upper right corner of Airport 1, the runway area of Airport 2, and the lower left corner of Airport 4, leading to relatively high false alarm rates across most algorithms. Although Airport 3 lacks prominent interfering objects, the complex variety of background elements and the relatively low spatial resolution of the dataset pose significant challenges for the algorithms. Based on the visualizations in Fig. \ref{fig4}, traditional methods, except for the LRASR, exhibit relatively lower false alarm rates for G-RX, L-RX, and CRD algorithms, but they also suffer from reduced accuracy, particularly in detecting targets in the Airport 3. On the other hand, deep learning-based approaches such as CNN, Auto-AD, and LRR-Net demonstrate significantly superior accuracy. Overall, among all algorithmic detection results, LRR-Net$^+$ outperforms others in noise interference suppression and showcases strong robustness.

Fig. \ref{fig5} shows the ROC curves, enabling a quantitative evaluation of the performance of all the contrast detectors. Notably, the performance of LRR-Net$^+$, indicated by the red curve, surpasses that of the other methods. Specifically, at a FAR value of 0.6, LRR-Net$^+$ achieves a PD value of 0.9 with a stable convergence, while the PD values of most other methods are around 0.8 or lower. Tab. \ref{Tab4} provides the AUC values for all comparative methods on the four Airport datasets, and the sixth row displays the average detection accuracy across these datasets. Overall, the Deep Learning-based methods, specifically CNN, Auto-AD, and LRR-Net, exhibit superior performance compared to the G-RX, CRD, and LRASR methods in general. Furthermore, L-RX demonstrates exceptional detection performance, surpassing CNN. These findings align with the outcomes obtained from the visual comparison.

\textbf{AIR-HAD-Vehicle:} In addition to the interference caused by trees and shadows in the background, the research dataset exhibits the following phenomena: Strong false alarm targets appear in the middle region of Fig. \ref{fig6} and the lower region of Fig. \ref{fig7}. These false alarm targets are artificial building structures, exhibiting similar brightness and shapes to the anomalous targets. In Fig. \ref{fig6}, apart from the four methods, CNN, Auto-AD, LRR-Net, and LRR-Net$^+$, the other methods failed to detect a vehicle with relatively low reflectance in the middle region, which is prone to confusion with the surrounding shadows. However, even so, CNN shows a significantly higher false negative rate, and Auto-AD and LRR-Net are unable to mitigate the false alarm rate caused by artificial buildings. In Fig. \ref{fig7}, the main issues arise with the first vehicle on the right and the first vehicle from the top, concerning false negatives, as well as the higher false alarm rate due to artificial buildings. Overall, among all the compared methods, LRR-Net$^+$ demonstrates the best performance, exhibiting excellent resistance to interference from trees, shadows, and artificial buildings, while also displaying remarkable vehicle detection capabilities.

Fig. \ref{fig9} (a) and (b) illustrate the ROC curves of LRR-Net$^+$ and other comparative methods on the AIR-HAD-Vehicle dataset. In contrast to the (a) graph, the brown curve in (b) representing LRR-Net$^+$ exhibits outstanding detection performance right from the beginning. However, in the (a) graph, the initial performance of the LRR-Net$^+$ curve appears relatively ordinary. Nevertheless, with an increase in the number of iterations, especially after reaching a False Alarm Rate (FAR) of 0.02, the brown curve shows a sharp rise, creating a noticeable gap compared to other algorithms, and it enters a plateau phase at FAR 0.05. Meanwhile, the performance of other algorithms continues to improve. In Tab. \ref{Tab4}, the seventh row demonstrates that, compared to the Airport dataset, the detection performance of the first four traditional models drops significantly, while LRR-Net$^+$ consistently maintains a relatively excellent performance on this dataset, confirming the previous statements.

\begin{figure*}[!t]
	  \centering
			\includegraphics[width=0.9 \textwidth]{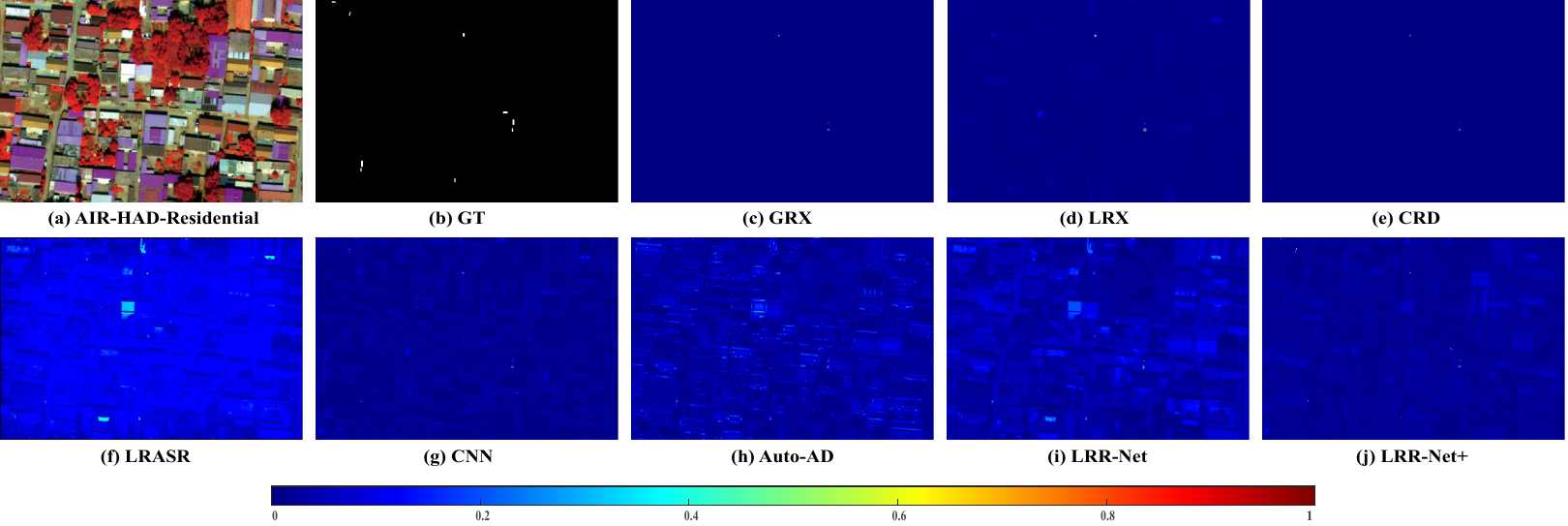}
        \caption{Image descriptions and detection maps of AIR-HAD-Resident, showing (a) false-color images, (b) corresponding ground truth, and the detection maps corresponding to (c)–(g) Global-RX, Local-RX, CRD, LRASR, CAE, Auto-AD, LRR-Net, and LRR-Net$^+$ respectively.}
\label{fig8}
\end{figure*}

\begin{figure*}[!t]
	  \centering
			\includegraphics[width=0.85\textwidth]{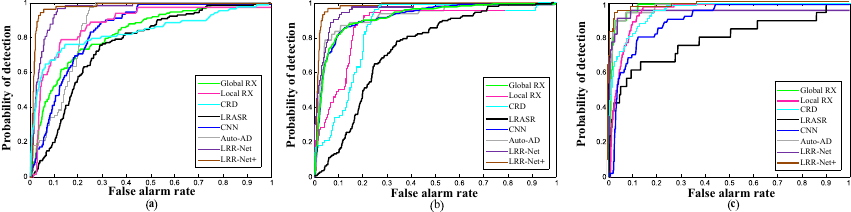}
        \caption{ROC curves obtained by different methods for AIR-HAD-Vehicle-I, AIR-HAD-Vehicle-II and AIR-HAD-Resident datasets.}
\label{fig9}
\end{figure*}

\textbf{AIR-Resident: } Concerning the visual assessment, Fig. \ref{fig8} visually presents the detection results of the eight methods applied to the residential dataset. The large scale of this dataset, the more complex scenes, and the more disturbing information undoubtedly pose a great challenge to the detection accuracy of the algorithm. The anomalous part refers to the nine cars parked in the building area. There is no doubt that the G-RX, LRX, CRD, and LRASR algorithms, none of them escaped the interference brought by the edges of the buildings. They can only identify the three more obvious anomalies in the middle. As can be seen from Tab. \ref{Tab4}, none of their detection results reach 0.75. In contrast, CNN and Auto-AD are slightly more resistant to interference, but still insensitive to the three anomalies in the upper left corner and the bottommost part. The anomalies detected by LRR-Net$^+$ exhibit significantly higher sensitivity in the scenarios of the AIR-HAD-Resident dataset compared to other methods.

In terms of evaluation and comparison of ROC curves, Fig. \ref{fig9} (c) presents a performance comparison of various methods on the AIR-HAD-Resident dataset. In the early iterations, the LRR-Net algorithm quickly achieves a stable solution through the optimization of fixed dictionary atoms, and its initial advantage is further enhanced by incorporating the advantages of deep learning. However, as the number of iterations increases, the LRR-Net$^+$ algorithm shows a significantly improved rise rate. This is because the alternating update of dictionary atoms and coefficients increases the search space of solutions, overcoming the limitation of fixed dictionary atoms' search range. Considering the AUC values in Tab. \ref{Tab4}, due to the large scale and high complexity of the AIR-HAD-Resident dataset, all algorithms produce relatively lower detection results compared to the aforementioned dataset. Overall, GRX, LRX, CRD, and LRASR face challenges in handling complex large-scale scenes, resulting in AUC values below 0.7. On the other hand, deep learning-based methods demonstrate certain advantages, but CNN exhibits a higher miss detection rate, while LRR-Net and Auto-AD have higher false alarm rates, and their AUC values cannot reach 0.8. In contrast, LRR-Net$^+$ performs relatively better, especially in detecting anomalies in the upper-left corner.

\subsection{Discussion}
In conventional physics-based approaches, the G-RX and L-RX algorithms are fundamentally based on utilizing the squared Mahalanobis distance as a statistical measure to assess the spatial dissimilarity between the target pixels and the global or local background pixels in the image data. These methods only focus on global or local information, and they lack sensitivity to anomalous targets in complex scenes. The CRD algorithm comprehensively leverages both global and local features by employing a sliding dual-window technique to approximate the value of each pixel. In this approach, the pixel values are modeled as a simple linear combination of neighboring samples, thereby achieving enhanced pixel value representation and context awareness within the image. However, this algorithm still has limited ability to differentiate between anomalous and background information. It is evident from the detection results that the LRASR model, which combines sparsity and low rankness, exhibits similarities to the aforementioned three methods. The detector models based on conventional methods manifest constrained adaptability, being tailored primarily for specific data types, and exhibit suboptimal sensitivity to interference information, including shadows, trees, and edges, in intricate scenes. Consequently, the detection results display conspicuous fluctuations, underscoring the challenges encountered in addressing diverse environmental complexities and variations.

In contrast, deep learning techniques including CNN and Auto-AD, which fall under the category of AE models, demonstrate remarkable data mining capabilities and exhibit commendable robustness in detecting anomalies across datasets with diverse complexities. Nonetheless, attributable to the dearth of fundamental prior knowledge, the search directions of these methods lack effective control, leading to significant fluctuations and a tendency to become trapped in local optima during optimization. Conversely, Our proposed LRR-Net$^+$ model synergistically integrates the strengths of both paradigms by leveraging the LRR model as a form of prior knowledge to guide and inform the network's optimization process. This incorporation of prior knowledge offers more structured and controlled optimization, resulting in superior performance consistently surpassing that of other methodologies. These findings underscore the potential of interpretable deep unfolding techniques in the HAD domain.

In summary, based on the detection plots generated by the aforementioned algorithms, it is evident that our LRR-Net$^+$ consistently outperforms the others across various sizes and types of anomalies, as well as different complexities of backgrounds. This observation serves as strong evidence of the exceptional robustness demonstrated by our proposed LRR-Net$^+$.

\subsection{Performance Analysis}
In this section, we aim to evaluate the efficacy of incorporating physical models into deep unfolding techniques within the context of HAD. This evaluation will be accomplished through a comprehensive comparative analysis involving multiple iterations of network and model optimization. To assess the performance of the model depicted in Figure \ref{fig8}, two key indicators, Mean Squared Error (MSE) and Precision-Recall Effectiveness (PRE), will be employed, which definitions are as follows

\begin{equation}
\label{eq23}
\begin{aligned}
\mathbf{MSE}=\frac{\lVert\mathbf{X}-\mathbf{R}_{\mathit{recon}}\rVert^2}{\lVert\mathbf{X}\rVert^2},
\end{aligned}
\end{equation}
where $\mathbf{X}$ denotes the original image and $\mathbf{R}_{\mathit{recon}}$ represents the recovered image, and 
\begin{equation}
\label{eq24}
\begin{aligned}
\mathbf{PRE}=\frac{\lVert\mathbf{R}_{\mathit{S}}-\mathbf{R}_{\mathit{gt}}\rVert^2}{\lVert\mathbf{R}_{\mathit{gt}}\rVert^2},
\end{aligned}
\end{equation}
where $\mathbf{R}_{\mathit{S}}$ denotes the anomaly image and $\mathbf{R}_{\mathit{gt}}$ represents the ground truth image.

\begin{figure}[!t]
	  \centering
	\includegraphics[width=0.8\textwidth]{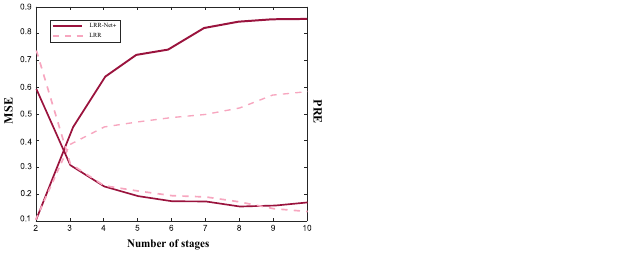}
        \caption{Achievable detection performance of LRR-Net$^+$ concerning several updating stages: (a) MSE, (b) PRE.}
\label{fig10}
\end{figure}

In Figure \ref{fig10}, our experimental findings indicate that when the number of iterations is below 15, both the traditional LRR model and LRR-Net$^+$ exhibit similar trends in anomaly detection accuracy. This similarity can be attributed to the fact that both models initialize their parameters using the ADMM algorithm. However, as the number of iterations increases, particularly within the range of 15 to 40 iterations, LRR-Net$^+$ experiences a sharp improvement in anomaly detection accuracy, while the LRR model shows a relatively slower enhancement. This behavior may be attributed to two underlying reasons: Firstly, the traditional LRR model only updates dictionary coefficients, which are constrained by the use of fixed atoms in the optimization process, thereby limiting the performance of dictionary learning. On the other hand, LRR-Net$^+$ capitalizes on the data mining capabilities of deep learning, transforming the manual adjustment process into a series of learnable parameters updated automatically by the network.
Furthermore, it is important to note that, in comparison to the conventional model, which often necessitates hundreds of iterations, the detection performance of LRR-Net$^+$ tends to stabilize after 40 iterations. This observation demonstrates that a mere 40 iterations based on the deep unfolding technique are adequate to achieve satisfactory anomaly detection performance. As such, LRR-Net$^+$ can effectively replace the traditional joint dictionary learning method, which typically requires a considerably larger number of iterations, without compromising the quality of detection results.

\section{Conclusion}\label{Conclusion}
Anomaly detection has long been considered one of the main research topics in many fields, such as images and signals. Hyperspectral remote sensing images have attracted wide attention from researchers due to their rich spectral bands. However, only from the existing low-resolution, simple background, small-scale HAD public data resources, limiting the performance of the model in the separation of background and target features and relying on manual parameter selection. To this end, we build a new set of HAD benchmark datasets for improving the robustness of the HAD algorithm in complex scenarios, AIR-HAD for short. Further, we propose a generalized and interpretable HAD network by deeply unfolding a dictionary-learnable LLR model, named LRR-Net$^+$, which is capable of spectrally decoupling the background structure and object properties in a more generalized fashion and eliminating the bias introduced by vital interference targets concurrently. In addition, LRR-Net$^+$ incorporates the solution process of the ADMM optimizer as prior knowledge within the deep network, facilitating parameter optimization with enhanced interpretability. Through rigorous comparisons with state-of-the-art HAD methods, LRR-Net$^+$ demonstrates superior anomaly detection performance across all seven datasets, outperforming its competitors. In our future endeavors, we aspire to further expand these datasets to a larger scale while simultaneously developing corresponding feature learning models. Specifically, we aim to leverage more potent deep prior techniques that embed additional interpretable knowledge or priors to guide the network optimization in the context of the HAD task. This pursuit seeks to enhance the efficiency and interpretability of the HAD algorithms and contribute to advancing state-of-the-art Abnormal detection technology.
\bibliographystyle{IEEEtran}
\bibliography{LRR-Net_ref}
\end{document}